\documentclass{article}

\def\DH{\rm I\kern-1.5pt\rm H\kern-1.5pt\rm I}

\def\DR{\rm I\kern-1.45pt\rm R}
\def\DC{\kern2pt {\hbox{\sqi I}}\kern-4.2pt\rm C}

\newcommand{\cF}{{\cal F}}
\newcommand{\cA}{{\cal A}}

\newcommand{\bD}{{\overline D}}
\newcommand{\bX}{{\overline X}}
\newcommand{\bW}{{\overline W}}

\newcommand{\tF}{{\widetilde F}}
\newcommand{\hF}{{\widehat F}}
\newcommand{\bxi}{{\bar\xi}}
\newcommand{\bpsi}{{\bar\psi}}
\newcommand{\tpsi}{{\tilde\psi}}

\newcommand{\bnu}{{\bar\nu}}
\newcommand{\tnu}{{\tilde\nu}}
\newcommand{\bm}{{\bar m}}

\newcommand{\nn}{\nonumber}
\newcommand{\ba}{\begin{array}}
\newcommand{\ea}{\end{array}}
\newcommand{\be}{\begin{equation}}
\newcommand{\ee}{\end{equation}}
\newcommand{\bea}{\begin{eqnarray}}
\newcommand{\eea}{\end{eqnarray}}
\newcommand{\bi}{\begin{itemize}}
\newcommand{\ei}{\end{itemize}}

\newcommand{\eps}{\varepsilon}

\newcommand{\p}[1]{(\ref{#1})}

\def\im{{\rm i}}

\begin{document}\thispagestyle{empty}

\begin{flushright}
%Draft \\
%Version 4.0 \\
%\today \\
\end{flushright}\vspace{2cm}

\begin{center}
{\Large\bf Testing the FPS approach in $d=1$}
\end{center}
\vspace{1cm}

\begin{center}
{\large\bf S.~Bellucci${}^a$, S.~Krivonos${}^{b}$,
A.~Sutulin${}^{a,b}$}
\end{center}

\begin{center}
${}^a$ {\it
INFN-Laboratori Nazionali di Frascati,
Via E. Fermi 40, 00044 Frascati, Italy} \vspace{0.2cm}

${}^b$ {\it
Bogoliubov  Laboratory of Theoretical Physics, JINR,
141980 Dubna, Russia} \vspace{0.2cm}

\end{center}
\vspace{2cm}

\begin{abstract}\noindent
We apply  the approach of S.~Ferrara, M.~Porrati and A.~Sagnotti \cite{FPS} to the one dimensional system described by  the $N=2, d=1$ supersymmetric
action for two particles in which one of $N=1$ supersymmetries is spontaneously broken.  Using the nonlinear realization approach we reconsider
the system in the basis where only one superfield has the Goldstone nature while the second superfield can be treated as the matter one, being invariant
under transformations of the spontaneously broken $N=1$ supersymmetry. We establish the transformations relating the two selected FPS-like cases with our
more general one, and find the field redefinitions which relate these two cases. Thus we demonstrate, at least in one dimension, that the only difference
between two FPS cases lies in the different choice of the actions, while the supermultiplets specified by the FPS-like constraints are really the same.
Going further with the nonlinear realization approach, we construct the most general action for the system of two $N=1$ superfields possessing one
additional hidden spontaneously broken $N=1$ supersymmetry. The constructed action contains two arbitrary functions and reduces to the FPS actions upon
specification of these functions. Unfortunately, the exact form of these functions corresponding to FPS actions is not very informative and gives no
explanation on why the FPS cases are selected.
\end{abstract}

\newpage
\setcounter{page}{1}
\setcounter{equation}{0}
\section{Introduction}

\setcounter{equation}{0}
 The key idea of Bagger and Galperin approach in constructing the $N=2, D=4$ supersymmetric Born-Infeld theory was to embed $N=1$ vector supermultiplet
 $W_\alpha$ along with $N=1$ chiral one $X$ into $N=2$ chiral supermultiplet $(X, W_\alpha)$ in such a way that the transformations of additional hidden $N=1$ supersymmetry
 with  parameters $\eta_{\alpha}, \bar\eta_{\dot\alpha}$
\be\label{i1}
\delta \left( W\right)_\alpha = \left( 1-\frac{1}{4} \bD{}^2 \bX \right)\eta_\alpha -\im \partial_{\alpha\dot\alpha} X \bar\eta{}^{\dot\alpha}, \;
\delta X = -2 \left( W \right)^\alpha \eta_\alpha ,
\ee
describe its spontaneous breaking due to the presence of the constant term in the transformation law of $W_\alpha$.
In view of the property of $N=1$ vector supermultiplet
$$ D^2 W_\alpha \sim \partial_{\alpha\dot\alpha} \bW^{\dot\alpha}$$
the trivial action
\be\label{i2}
S= \int d^4x d^2 \theta X
\ee
becomes invariant under transformations \p{i1}. The action \p{i2} acquires a real meaning after introducing the additional constraint,
invariant under \p{i1} transformations
\be\label{i3}
W\cdot W +X \left( m -\frac{1}{4} \bD{}^2 \bX \right) =0,
\ee
which can be solved in order to express $X$ in terms of the superfields $W_\alpha, \bW_{\dot\alpha}$ \cite{BG}. The bosonic core of the resulting action
is just the Born-Infeld action \cite{BG}. Thus, the $N=2$ supersymmetric Born-Infeld theory describes the partial spontaneous breaking of $N=2$
supersymmetry to $N=1$ one with a $N=1$ vector supermultiplet as the corresponding Goldstone supermultiplet.

In the  recent  paper \cite{FPS} S.~Ferrara, M.~Porrati and A.~Sagnotti proposed the generalization of J.~Bagger and A.~Galperin construction
to the cases of several $N=1$ vector supermultiplets. The FPS approach includes two basic ingredients. First, the nonlinear constraint \p{i3}
is generalized to be
\be\label{i4}
d_{abc} \left( W_b \cdot W_c +X_b \left( m_c -\frac{1}{4} \bD{}^2 \bX_c \right)\right) =0,
\ee
where $W_{a \alpha}, X_a$ are $n$-copies of $N=1$ vector multiplet, $d_{abc}$ are totally symmetric constants and $m_a$ is a set
of constants. The realization of hidden $N=1$ supersymmetry is the same as in the Bagger and Galperin case
\be\label{i5}
\delta  W_{a \alpha} = \left( m_a -\frac{1}{4} \bD{}^2 \bX_a \right)\eta_\alpha -\im \partial_{\alpha\dot\alpha} X_a \bar\eta{}^{\dot\alpha}, \;
\delta X_a = -2 W_a^\alpha \eta_\alpha .
\ee
The invariance of the constraint \p{i4} with respect to the hidden $N=1$ supersymmetry transformations \p{i5}
introduced the additional constraint (which is trivially satisfied in the case of one supermultiplet)
\be\label{i6}
d_{abc} W_{b \alpha}  X_c =0 .
\ee
The constraints \p{i4}, \p{i6} can be also solved expressing the bosonic $N=1$ chiral superfields $X_a$ in terms of the $N=1$ vector supermultiplets $W_{a \alpha}$.

The next nontrivial step of FPS approach is the structure of the corresponding action
\be\label{i7}
S= \int d^4x d^2 \theta \;\left[  e^a X_a+ C_{ab} \left( W_a \cdot W_b +X_a \left( m_b -\frac{1}{4} \bD{}^2 \bX_b \right)\right)+ c.c.\right],
\qquad  e^a = const, C_{ab}=C_{ba}=const.
\ee
The additional term with $C_{ab}$ is also invariant with respect to transformations \p{i5}. Such a term does not exist
in the case of one supermultiplet, but it proved to be essential in the cases of several supermultiplets \cite{FPS}
\footnote{We are thankful to S.~Ferrara, M.~Porrati and A.~Sagnotti for the correspondence concerning the essentiality of this term.}.
The action \p{i7} with $X_a$ being the solution of the constraints \p{i4}, \p{i6} is treated in \cite{FPS} as the many-field
extension of $N=2$ Born-Infeld theory.

In paper \cite{FPS} the detailed analysis of $n=2$ case was presented, which can be divided into two subcases:
\begin{itemize}
\item $d_{111}=1, d_{112}=-1 \quad I_4=0$,
\item $d_{111}=1, d_{122}=-1 \quad I_4>0$,
\end{itemize}
where $I_4$ is a quartic invariant, discussed in \cite{FPS} (see also the refs. therein).
Then the analysis of $n=3$ case has been performed in \cite{FPS2}.

The above, quite short sketch of the FPS approach is enough to raise two interesting questions
\begin{itemize}
\item whether the supermultiplets with different constants $d_{abc}$ are really different?
\item whether the action \p{i7} is unique as it happened in $n=1$ case?
\end{itemize}
The reasons for raising the first question are the following. After resolving the constraints FPS \p{i4}, \p{i6}
we will have in the theory $n$ copies of $N=1$ vector multiplets $W_{a \alpha}$ with highly nonlinear transformation properties
under hidden $N=1$ supersymmetry \p{i5}. The presence of the constants $m_a$ in \p{i5} at first sight means that
all of them have Goldstone nature. But it cannot be so, because the spontaneous breaking of $N=2$ to $N=1$ supersymmetries accompanied by the
presence in the theory of only one Goldstone superfield \cite{NLR1,NLR2}. Moreover, as it follows from an alternative construction of $N=2$
supersymmetric Born-Infeld action within the nonlinear realization approach discussed in \cite{BG}, it should exist a special basis in which
only one superfield is Goldstone superfield, while the remaining superfields have to be matter superfields with the  trivial transformation
properties with respect to broken supersymmetry. In view of the relation between linear and nonlinear realizations of partially broken
supersymmetries \cite{ikap1}, one should expect that any linear realization of broken supersymmetry can be transformed by a proper
highly nonlinear, but invertible field redefinition, to the {\it same} nonlinear realization in this basis. Of course, the nonlinear
realization approach is less useful for constructing of superfield actions, but the claim that different choices of the constants $d_{abc}$
in the basic constraints \p{i4}, \p{i6} must be equivalent, should be carefully analyzed.

The second question about uniqueness of the action we are raising is also related with existence of the special basis in which we have one
Goldstone and an arbitrary number of matter superfields. In such a basis one may easily construct the general action
which will have a functional freedom. Thus, it has to be quite interesting to analyze this general action and understand
the reasons that just select the FPS actions.

Unfortunately, the analysis of $N=2$ extended Born-Infeld theory is quite involved, and the ideological part is hidden
behind complicated calculations. Fortunately enough the FPS approach, as we already stressed in \cite{comments},
is not limited by application to Born-Infeld theory only. Instead, it will work perfectly for many other systems with partial  breaking of
supersymmetry. The simplest of all such examples is provided by a theory describing partial breaking of $N=2$ to $N=1$ supersymmetry in
one dimension - the textbook case considered in \cite{DIK}. The FPS
generalization of such a system, being quite simple, will inherit all essential properties of the FPS construction.
In the present paper we are exploring the generalization of FPS construction for this case. In the next Section
we remind the description of $N=2 \rightarrow N=1$ partial breaking of supersymmetry in $d=1$ \cite{DIK} and
then extend it to the many-particle case following to FPS approach. In Section 3 we explicitly establish the transformations that relate two
different $n=2$ cases with the special basis discussed above, as well as
the explicit relations between them. In Section 4 we construct the general action for such a system containing
two arbitrary functions and provide the explicit form of these functions reducing the general action to FPS-like ones.
Finally, in Conclusion we summarize our results.
\setcounter{equation}{0}
\section{Partial breaking in one dimension: superparticle action and its generalization}
Let us start, following~\cite{DIK}, with a simple example of superparticle model describing $N=2\; \rightarrow\; N=1$ partial
breaking of global supersymmetry in $d=1$.
In this case, the anticommutation relation of $N=1, d=1$ Poincar\'e superalgebra reads
\be\label{alg-N1}
\big \{Q, Q \big \} = 2P\,.
\ee
The coordinates $(t, \theta)$ of $N=1$ superspace satisfy the following rules of complex conjugation:
$t^{\dagger} = -t, \theta^{\dagger} = \theta$.
Next, we define the bosonic and fermionic superfields $v(t, \theta)$ and $\psi(t, \theta)$ related as
\be\label{fields}
\psi = \frac{1}{2}\, Dv, \quad (v^{\dagger} = -v,\;  \psi^{\dagger} = \psi),
\ee
where the spinor derivative $D$ obeys the relation\footnote{The time derivative of any variable $f$ denoted as: $\dot f= \partial_t f$.}
\be\label{der1}
D = \frac{\partial}{\partial \theta} + \theta \partial_t, \quad
\big \{D, D \big \} = 2 \partial_t\,.
\ee
\subsection{Linear realization: one particle case}
To realize an additional spontaneously broken $N=1$ supersymmetry we introduce, in full analogy with \cite{BG}, two spinor superfields
$\psi$ and $\nu$ transforming under the $S$-supersymmetry as
\be\label{S-susy}
\delta \psi = \eps (1 - D \nu)\,, \quad \delta \nu = \eps D \psi.
\ee
The presence of the constant shift in \p{S-susy} suggests the interpretation of $\psi$ as the Goldstone fermion accompanying  the $N=2 \rightarrow N=1$ breaking.
The superfield $\nu$, due to its transformation properties under the $S$-supersymmetry \p{S-susy}, may be chosen as a Lagrangian density, since the integral
\be\label{action0}
S = \int dt d\theta\, \nu
\ee
is invariant with respect to both broken and unbroken supersymmetries. To be meaningful, the action \p{action0} should be accompanied
an additional constraint, which is invariant under transformations \p{S-susy} and which, by analogy with \p{i3}, will allow to express
the superfield $\nu$ through the superfield $\psi$. One may easily check, that the corresponding invariant constraint
looks quite similar to \p{i3} to be \cite{DIK}
\be\label{c1}
\psi D \psi - \nu \left( 1 - D \nu\right) =0
\ee
with evident solution given by
\be\label{c2}
\nu = \frac{2 \psi D \psi}{1+\sqrt{1-4 (D\psi)^2}}\,.
\ee
Note, that from \p{c1} follows
\be\label{cc1}
\psi \nu =0.
\ee
The bosonic core of the action \p{action0} with the integrand \p{c2} corresponds to the action of particle in $d=1$
\be\label{c3}
S_{bos} = \frac{1}{2} \int dt \left( 1 - \sqrt{1- (\dot v)^2}\right).
\ee
Thus, this simplest system demonstrate a close analogy with the Bagger and Galperin construction of the supersymmetric Born-Infeld action.
\subsection{Generalization to the multiparticle case}
The above considered case can be easily extended to that of an arbitrary number of $N=1$ superfields strictly following the FPS
approach. Doing so, we firstly introduce a set of $n$ superfields $\psi_a$ and $\nu_a$ transforming under implicit $N=1$
supersymmetry as
\be\label{FPS-susy}
\delta \psi_a = \eps (m_a - D \nu_a)\,, \quad \delta \nu_a = \eps D \psi_a,
\ee
where $m_a$ are arbitrary constants. Now, again in full analogy with FPS approach, we impose the following generalization of the constraints \p{c1}
\be\label{FPS-constr}
d_{abc} \left(  \psi_b D \psi_c - \nu_b \left (m_c- D \nu_c \right )\right) = 0.
\ee
The invariance of these constraints under broken supersymmetry \p{FPS-susy} leads to an additional restriction
\be\label{add-c}
\partial_t \left(d_{abc} \psi_b \nu_c \right) = 0,
\ee
which can be reformulated as
\be\label{add-constr}
 d_{abc} \psi_b \nu_c = 0,
\ee
assuming that the constant of integration is equal to zero.
Finally,  the FPS-like generalization of the action \p{action0} reads
\be\label{action}
S = \int dt d\theta\,\left[  e^a \nu_a+ C_{ab}  \left(  \psi_a D \psi_b - \nu_a \left (m_b- D \nu_b \right )\right)\right] ,
\ee
where $e^a$ and $C_{ab}=C_{ba}$ are arbitrary real constants.
\setcounter{equation}{0}
\section{Standard basis for $n=2$ cases}
Up to now we have just repeated the basic steps of the FPS approach applying it to the one dimensional case.
Let us now consider in detail a system of two superparticles.

In the standard approach of the spontaneous breaking of  supersymmetry the two particle model should contain only one
Goldstone superfield (with the Goldstone fermion among their components). The rest of the fields must be matter ones, i.e. they
should not transform with respect to broken supersymmetry. Let us demonstrate how such a splitting works in the present system.
\subsection{The case with $d_{111}=d_{222}=1$}
With such a choice of the symmetric tensor $d_{abc}$, the basic constraints \p{FPS-constr}, \p{add-constr} have a splitting form
\be\label{constr1}
\left\{ \begin{array}{lcr}
\psi_1 D\psi_1 -\nu_1\left(m_1 - D \nu_1\right) =0,& \quad & \psi_1 \nu_1 =0, \\
\psi_2 D\psi_2 -\nu_2\left(m_2 - D \nu_2\right) =0,& \quad & \psi_2 \nu_2 =0.
\end{array} \right.
\ee
By rescaling of the variables $\psi_{a}$ and $\nu_{a}$  in \p{constr1}, one may always choose
\be\label{m1}
m_1=m_2=1.
\ee
The transformation properties of $\psi_{a}$ and $\nu_{a}$ \p{FPS-susy} with such a choice of parameters $m_{a}$ are
\be\label{tr1}
\left\{ \begin{array}{lcr}
\delta \psi_1 = \epsilon \left(1 - D \nu_1\right),& \quad & \delta\nu_1 =\epsilon D\psi_1 , \\
\delta \psi_2 = \epsilon \left(1 - D \nu_2\right),& \quad & \delta\nu_2 =\epsilon D\psi_2 .
\end{array} \right.
\ee
Clearly, the solution of the constraints \p{constr1} is quite similar to the one particle case \p{c2} and now it reads
\be\label{cc2}
\nu_1 = \frac{2 \psi_1 D \psi_1}{1+\sqrt{1-4 (D\psi_1)^2}}\,, \qquad \nu_2 = \frac{2 \psi_2 D \psi_2}{1+\sqrt{1-4 (D\psi_2)^2}}\,.
\ee
Let us now introduce {\it two} Goldstone spinor superfields $\xi_{a}$ with the following transformation properties
\be\label{xi1}
\delta \xi_1 = \epsilon + \epsilon\, \xi_1 \, \partial_t\, \xi_1, \qquad \delta \xi_2 = \epsilon + \epsilon\, \xi_2 \, \partial_t\, \xi_2.
\ee
It is easy to check that the tilded fields $\tpsi_{a}$ and $\tnu_{a}$ defined as
\be\label{tpsi1}
\tpsi_a = \psi_a -\xi_a \left(1- D\nu_a\right),\quad \tnu_a =\nu_a - \xi_a D\psi_a,
\ee
transform under \p{tr1}, \p{xi1} as (see, e.g. \cite{DIK})
\be\label{tr2}
\delta \tpsi_a =\epsilon\, \xi_a\, \partial_t\, \tpsi_a, \qquad \delta \tnu_a = \epsilon\, \xi_a \,\partial_t\, \tnu_a .
\ee
Thus, it is a covariant operation to put these superfields equal to zero
\be\label{constr11}
\left\{
\begin{array}{lcl}
\tpsi_a=0, & \qquad & \psi_a-\xi_a\left(1-D \nu_a\right)=0, \\
& \Rightarrow & \\
\tnu_a=0 &\qquad & \nu_a-\xi_a D \psi_a=0.
\end{array}
\right.
\ee
Moreover, the constraints \p{constr1} immediately follow from \p{constr11}. One should note that  the constraints $\tpsi_a=\tnu_a=0$ contain some
additional information: they can be used to express the superfields of the linear realization
$\psi_a$ and $\nu_a$ in terms of the Goldstone superfields $\xi_a$
\be\label{rel1}
\nu_a = \frac{\xi_a D \xi_a}{1+D\xi_a D \xi_a},\quad \psi_a =\frac{\xi_a}{1+D\xi_a D\xi_a} \qquad \qquad \mbox{(no summation over {\it a})}.
\ee
Until now we have two Goldstone superfields $\xi_{a}$, while we are expecting to have only one essential Goldstone fermionic
superfield and one matter superfield. This may be achieved by passing to the new superfields $\eta$ and $\lambda$
\be\label{eta1}
\eta= \frac{1}{2}\left(\xi_1+\xi_2\right) +\frac{1}{4} \xi_1 \xi_2 \left( \dot\xi{}_1 -\dot \xi{}_2\right), \quad
\lambda = \frac{1}{2}\left(\xi_1-\xi_2\right) +\frac{1}{4} \xi_1 \xi_2 \left( \dot \xi{}_1 +\dot \xi{}_2\right),
\ee
which, in virtue of \p{xi1}, transform as expected\footnote{This is the form-variation of the fields under implicit $N=1$ supersymmetry: $\delta \cA= \cA'(t,\theta)-
\cA(t,\theta)$.}
\be\label{eta2}
\delta \eta = \epsilon+\epsilon \eta \partial_t \eta, \qquad
\delta \lambda = \epsilon \eta \partial_t \lambda .
\ee
From \p{eta2} one concludes that $\eta$ is the fermionic Goldstone superfield accompanying the present spontaneous breaking of $N=2$ supersymmetry, while
$\lambda$ is a matter fermionic superfield. The superfields $\xi_a$ as well as $\psi_a, \nu_a$ can be expressed in the
terms of $\eta$ and $\lambda$:
\be\label{xi11}
\xi_1 = \eta + \lambda +\eta \lambda \left( \dot \eta + \dot \lambda \right), \quad
\xi_2 = \eta - \lambda -\eta \lambda \left( \dot \eta - \dot \lambda \right),
\ee
and
\bea\label{fin1}
&&\psi_1 =\frac{\eta+\lambda}{1+\left( D\eta+D\lambda\right)^2}+ \eta\lambda \left(\dot \eta+\dot \lambda\right)
\frac{1- \left( D\eta+D\lambda\right)^2}{\left[ 1+\left( D\eta+D\lambda\right)^2\right]^2},\quad
\psi_2 = \left.\psi_1\right|_{\lambda \rightarrow -\lambda}, \nn \\
&& \nu_1 =\left( D\eta+D\lambda\right)\left[ \frac{\eta+\lambda}{1+\left( D\eta+D\lambda\right)^2}+
\frac{2\eta\lambda \left(\dot \eta+\dot \lambda\right)}{\left[ 1+\left( D\eta+D\lambda\right)^2\right]^2}\right], \quad
\nu_2 = \left.\nu_1\right|_{\lambda \rightarrow -\lambda}.
\eea
Thus, the system of constraints \p{constr1} represents a non-standard description of the essential Goldstone superfield
$\eta$ and a matter fermionic superfield $\lambda$.

Clearly, the FPS-like action \p{i7} for this case reads
\be\label{action-tilde1}
S = \int dt d\theta \, \Big\{ e_1\,\nu_1 +e_2 \nu_2\, +C_{12}\big( \psi_1 D \psi_2 - \nu_1 (1 - D \nu_2)+\psi_2 D \psi_1 - \nu_2 (1 - D \nu_1)\big) \Big\}
\ee
and it is invariant with respect to both supersymmetries.
\subsection{The case with $d_{111}=1, d_{122}=-1$}
With this choice of constants  the constraints \p{FPS-constr} read
\bea\label{d-FPS-constr}
&&
\psi_1 D \psi_1 - \psi_2 D \psi_2 - \nu_1 \left(m_1 - D \nu_1\right) + \nu_2 \left(m_2 - D \nu_2 \right) = 0, \\
&&
\psi_1 D \psi_2 + \psi_2 D \psi_1 - \nu_1 \left(m_2 - D \nu_2\right) - \nu_2 \left(m_1 - D \nu_1 \right) = 0, \nn
\eea
while those in \p{add-constr} take form
\be\label{d-add-constr}
\psi_1 \nu_1 - \psi_2 \nu_2 = 0, \qquad \psi_1 \nu_2 + \psi_2 \nu_1 = 0.
\ee
One can introduce the complex superfields $\psi, \nu$ and complex parameter $m$ as
\bea\label{new-basis0}
&&
\psi = \psi_1 + \im \psi_2\,, \qquad \nu = \nu_1 + \im \nu_2\,, \qquad m = m_1 + \im m_2\,,\\
&&
\bpsi = \psi_1 - \im \psi_2\,, \qquad \bnu = \nu_1 - \im \nu_2\,, \qquad \bm = m_1 - \im m_2\,, \nn
\eea
and then rewrite eqs. \p{d-FPS-constr} and \p{d-add-constr} in the splitting form \cite{comments}
\bea\label{new-basis}
&&
\psi D \psi - \nu \left( m - D \nu \right) = 0\,, \quad
\psi \, \nu = 0\,, \\
&&
\bar \psi D \bar\psi - \bar\nu \left( \bar m - D \bar\nu \right) = 0\,, \quad
\bar\psi \, \bar\nu = 0\,. \nn
\eea
The equations \p{new-basis} have the evident solution
\be\label{sol11}
\nu = \frac{2 \psi D \psi}{m+ \sqrt{m^2 -4 (D \psi)^2}}\,, \quad \bar\nu = \frac{2 \bar\psi D \bar\psi}{\bar m+ \sqrt{\bar m^2 -4 (D \bar \psi)^2}}\,.
\ee
Finally, the FPS-like action \p{i7} reads
\be\label{action-tilde}
S = \int dt d\theta\, \Big\{ e\,\nu +e^* \bnu\, +C_{12}\big( \psi D \bpsi - \nu (\bm - D \bnu)+\bpsi D \psi - \bnu (m - D \nu)\big)\Big\}\,,
\ee
where $e=\frac{1}{2}\,(e_1- \im e_2),\; e^*=\frac{1}{2}\,(e_1+ \im e_2)$.
This action is also invariant with respect to both supersymmetries.

One should note, that one may always choose the constants $m_1, m_2$ as
\be\label{m}
m_1=1, \; m_2 = 0, \qquad \Rightarrow \qquad m=\bm=1.
\ee
The choice $m_2=0$ is achieved by passing to the fields $\rho, \mu$
$$
\psi_2 \rightarrow \rho = m_2 \psi_1 - m_1 \psi_2, \quad \mu = m_2 \nu_1 - m_1 \nu_2,
$$
and, as it follows from \p{FPS-susy}
\be\label{FPS-susy1}
\delta \rho = -\eps \, D \mu \,, \quad \delta \mu = \eps D \rho.
\ee
The choice $m_1=1$ is guaranteed by the rescaling in \p{FPS-susy}
$$
\psi_1 \rightarrow \tpsi =m_1 \psi_1, \quad \nu_1 \rightarrow \tnu =  m_1 \nu_1.
$$

Now we have a full analogy with the previously considered splitting case. Indeed, the complex superfields $\psi$ and $\nu$ transform with respect to broken supersymmetry as
\be\label{tr22}
\left\{ \begin{array}{lcr}
\delta \psi = \epsilon \left(1 - D \nu\right), \quad  \delta\nu =\epsilon D\psi , \\
\delta \bpsi = \epsilon \left(1 - D \bnu \right), \quad  \delta\bnu =\epsilon D\bpsi .
\end{array} \right.
\ee
and obey the constraints
\be\label{new-basis000}
\left\{ \begin{array}{lcr}
\psi D \psi - \nu \left( 1 - D \nu \right) = 0\,, \quad \psi \, \nu = 0\, \\
\bpsi D \bpsi - \bnu \left( 1 - D \bnu \right) = 0\,, \quad \bpsi \, \bnu = 0\,
\end{array} \right.
\ee
If we now introduce {\it two} Goldstone spinor superfields $\xi, \bxi$ with the following transformation properties
\be\label{xi22}
\delta \xi = \epsilon + \epsilon\, \xi \, \partial_t\, \xi, \qquad \delta \bxi = \epsilon + \epsilon\, \bxi \, \partial_t\, \bxi,
\ee
than the tilded fields $\tpsi, \bar\tpsi$ and $\tnu,\bar\tnu$, which are defined as
\be\label{tpsi22}
\tpsi = \psi -\xi \left(1- D\nu\right),\quad \tnu =\nu - \xi D\psi, \qquad
\bar\tpsi = \bpsi -\bxi \left(1- D\bnu\right),\quad \bar\tnu =\bnu - \bxi D\bpsi,
\ee
transform under \p{tr22}, \p{xi22} as
\be\label{tr23}
\delta \tpsi =\epsilon\, \xi\, \partial_t\, \tpsi, \qquad \delta \tnu = \epsilon\, \xi \,\partial_t\, \tnu, \qquad
\delta \bar\tpsi =\epsilon\, \bxi\, \partial_t\, \bar\tpsi, \qquad \delta \bar\tnu = \epsilon\, \bxi \,\partial_t\, \bar\tnu.
\ee
Thus, again there is a covariant operation to put these superfields equal to zero
\be\label{constr22}
\left\{
\begin{array}{lcl}
\tpsi=0,\;\; \tnu=0,  & \qquad & \psi-\xi\left(1-D \nu\right)=0,\;\; \nu-\xi D \psi=0, \\
& \Rightarrow & \\
\bar\tpsi=0,\;\; \bar\tnu=0  & \qquad & \bpsi-\bxi\left(1-D \bnu\right)=0,\;\; \bnu-\bxi D \bpsi=0.
\end{array}
\right.
\ee
Clearly, the constraints \p{new-basis000} immediately follow from \p{constr22}. The constraints \p{constr22} can be easily solved
\be\label{rel22}
\nu = \frac{\xi D \xi}{1+D\xi D \xi}\,,\quad \psi =\frac{\xi}{1+D\xi D\xi}\,, \qquad
\bnu = \frac{\bxi D \bxi}{1+D\bxi D \bxi}\,,\quad \bpsi =\frac{\bxi}{1+D\bxi D\bxi}\,.
\ee
Repeating all calculations from previous
subsection we will get that the superfields $\eta$ and $\lambda$ defined as
\be\label{eta21}
\eta= \frac{1}{2}\left(\xi+\bxi\right) +\frac{1}{4} \xi \bxi \left( \dot \xi -\dot \bxi\right), \quad
\lambda = \frac{\im}{2}\left(\xi-\bxi\right) +\frac{\im}{4} \xi \bxi \left( \dot \xi + \dot \bxi\right),
\ee
have the same transformation properties as in \p{eta2}
$$
\delta \eta = \epsilon+\epsilon \eta \partial_t \eta, \qquad
\delta \lambda = \epsilon \eta \partial_t \lambda .
$$
It is important that the transformations \p{eta21} are invertible and the superfields $\xi,\bxi$ may be expressed in terms of
$\eta, \lambda$ as
\be\label{inverse}
\xi = \eta - \im \lambda -\im \eta\lambda \left( \dot \eta -\im \dot \lambda\right), \qquad
\bxi = \eta + \im \lambda +\im \eta\lambda \left( \dot \eta +\im \dot \lambda\right).
\ee
Finally, the genuine superfields $\psi,\nu$ can be expressed in terms of the Goldstone
superfield $\eta$ and  matter superfield $\lambda$:
\bea\label{fin2}
&&\psi =\frac{\eta - \im \lambda}{1+\left( D\eta- \im D\lambda\right)^2}- \im \eta\lambda \left(\dot \eta- \im \dot \lambda\right)
\frac{1- \left( D\eta- \im D\lambda\right)^2}{\left[ 1+\left( D\eta- \im D\lambda\right)^2\right]^2},\quad
\bpsi = \left(\psi\right)^\dagger, \nn \\
&& \nu =\left( D\eta- \im D\lambda\right)\left[ \frac{\eta- \im \lambda}{1+\left( D\eta- \im D\lambda\right)^2}- \im
\frac{2\eta\lambda \left(\dot \eta- \im \dot \lambda\right)}{\left[ 1+\left( D\eta- \im D\lambda\right)^2\right]^2}\right], \quad
\bnu = \left(\nu\right)^\dagger.
\eea
\subsection{Relations between two cases}
Having at hands the {\it invertible} relations between $\xi_1,\xi_2$ and $\eta, \lambda$ \p{xi11} and between
$\eta, \lambda$ and $\xi,\bxi$ \p{eta21}, one may easily find the relations between $\xi,\bxi$ and $\xi_1,\xi_2$
\bea\label{twocases}
&&
\xi= \frac{1}{2}\left(1-\im\right) \xi_1+ \frac{1}{2}\left( 1+\im\right) \xi_2+ \frac{1}{2}\xi_1 \xi_2 \left(
\dot{\xi}_1 - \dot{\xi}_2\right), \;
\bar\xi= \frac{1}{2}\left(1+\im\right) \xi_1+ \frac{1}{2}\left( 1-\im\right) \xi_2+ \frac{1}{2}\xi_1 \xi_2 \left(
\dot{\xi}_1 - \dot{\xi}_2\right), \nn\\
&&
\xi_1= \frac{1}{2}\left(1+\im\right) \xi + \frac{1}{2}\left( 1-\im\right) \bar\xi + \frac{1}{2}\xi \bar\xi \left(
\dot{\xi} - \dot{\bar\xi}\right), \;
\xi_2= \frac{1}{2}\left(1-\im\right) \xi + \frac{1}{2}\left( 1+\im\right) \bar\xi + \frac{1}{2}\xi \bar\xi \left(
\dot{\xi} - \dot{\bar\xi}\right) .
\eea
Thus, at least in one dimension, the two cases are completely equivalent, because they are related by {\it invertible}
fields redefinition \p{twocases}. The only difference between these cases lies in the different definition of the invariant actions \p{action-tilde}.
Moreover, these two actions, being clearly related in view of \p{twocases},
are not the most general ones. In the next Section we will construct the most general action for the Goldstone fermion
$\eta$ and the matter fermionic superfield $\lambda$.

\setcounter{equation}{0}
\section{Nonlinear realization approach}
An alternative description of the discussed system with partially broken $N=2,d=1$ supersymmetry may be provided
by the nonlinear realizations approach. It turns out that in the present case this approach is more suitable for the construction of
the most general superfield action. Let us demonstrate how such an action can be derived.
\subsection{Key ingredients}
We start with the $N=2, d=1$ Poincar\'{e} superalgebra with one central charge generator $Z$
\be\label{algebra}
\big \{ Q,Q \big\} = 2P, \quad \big \{ S,S \big\} = 2P, \quad \big \{ Q,S \big\} = 2Z.
\ee
Introducing a coset element $g$ as
\be\label{coset}
g = e^{tP} e^{\theta Q} e^{q Z} e^{\eta S},
\ee
and calculating the expression $g^{-1} dg = \omega_P P + \omega_Q Q + \omega_Z Z + \omega_S S$,
one finds the explicit expressions for the Cartan forms
\be\label{Cartan}
\omega_P = dt - d \theta \theta - d \eta \eta, \quad  \omega_Q = d \theta, \quad
\omega_Z = d q - 2 d \theta \eta, \quad \omega_S = d \eta.
\ee
The covariant derivatives can be found in a standard way, and in the present case they read
\be\label{der}
\nabla_{\theta} = D + \eta D \eta \nabla_t, \quad
\nabla_t = E^{-1} \partial_t,
\ee
where
\be\label{DE}
E = 1+ \eta \partial_t \eta, \quad E^{-1} = 1 - \eta \nabla_t \eta.
\ee
These derivatives satisfy the following (anti)commutation relations
\be\label{alg-der}
\big \{ \nabla_{\theta}, \nabla_{\theta} \big \} = 2 \left( 1 +  \nabla_{\theta} \eta \nabla_{\theta} \eta\right) \nabla_t, \quad
\big [ \nabla_t, \nabla_{\theta} \big ] = 2 \nabla_t \eta \nabla_{\theta} \eta \nabla_t.
\ee
Acting on \p{coset} from the left by different elements of the $N=2, d=1$ Poincar\'{e} supergroup $g_0$ one can find the transformation properties
of the coordinates $(t, \theta)$ and the Goldstone superfield $\eta(t, \theta)$. In particular, under both unbroken and broken supersymmetries they transform as follows:
\begin{itemize}
\item Unbroken supersymmetry ($g_0=e^{\epsilon Q}$\,):
\be\label{susyQ}
\delta_Q t = - \epsilon \theta, \quad
\delta_Q \theta=\epsilon,
\ee
\item Broken supersymmetry ($g_0=e^{\eps S}$\,):
\be\label{susyS}
\delta_S t = - \eps \eta, \quad \delta_S q = -2 \eps \theta, \quad
\delta_S \eta = \eps.
\ee
\end{itemize}
The final step is to express the superfield $\eta$ in terms of the bosonic superfield $q$ by imposing the constraint \cite{ih}
\be\label{ih}
\omega_Z|_{d\theta} =0 \qquad \Rightarrow \qquad \eta=\frac{1}{2}\nabla_{\theta} q.
\ee
It should be noted that in order to describe the matter superfield $\lambda$ within the nonlinear realization approach,
one should postulate its transformation properties under $S$ supersymmetry to be\footnote{Due to the transformation of the time $t$ in \p{susyS} the form-variation of $\lambda$  will be $\delta \lambda = \eps \eta \partial_t \lambda$.}
\be\label{sl}
\delta_S \lambda =0.
\ee
One may also define the bosonic superfield $\phi$ with the transformation property
\be\label{sphi}
\delta_S \phi =0,
\ee
which is related with $\lambda$ in the same manner as in \p{ih}
\be\label{ih2}
\lambda = \frac{1}{2} \nabla_{\theta} \phi.
\ee

\subsection{General action}

The most general Ansatz for the $N=1$ superfield action describing the Goldstone fermionic superfield $\eta$ and
the matter fermionic superfield $\lambda$ and having no dimensional constants reads
\be\label{S1}
S= \int dt d\theta \left( 1+ \eta \dot \eta\right) \left[ \eta\; F_1 + \lambda\; F_2+ \eta \dot \eta \lambda\; F_3 +
\eta \lambda \nabla_t \lambda\; F_4 \right],
\ee
where $F_1, F_2, F_3, F_4$ are arbitrary functions depending on $\nabla_{\theta}\eta$ and $\nabla_{\theta}\lambda$, only. The reason for such a
form of Ansatz is quite understandable:
\begin{itemize}
\item the functions $F_1, F_2, F_3, F_4$ have to be defined by imposing the invariance of the action \p{S1} with respect
to broken $S$-supersymmetry
\item with respect to $S$ supersymmetry, $\delta_S t =- \eps \eta,\; \delta_S \eta = \eps,\; \delta_S \lambda=0$,
the ``improved'' measure $dt d\theta \left( 1+ \eta \dot \eta\right)$ is invariant
\item the functions $F_1, F_2, F_3, F_4$ as well as $\nabla_t \lambda$ are also invariant with respect to $S$-supersymmetry.
\end{itemize}
Therefore, the variation of the action \p{S1} with respect to broken supersymmetry has a very simple form
\be\label{S2}
\delta_S S =  \int dt d\theta \;\eps \left[ \left( 1 + \eta \dot \eta\right) F_1 + \dot \eta \lambda F_3 + \lambda \dot \lambda F_4 \right].
\ee
Thus, the function $F_2$ may be chosen to be an arbitrary function of its arguments. To fix the functions $F_1,F_3$ and $F_4$
from the condition $\delta_S S=0$ one has, firstly, to integrate\footnote{The problem is that the integrand in \p{S2} can
be represented as a $D$ derivative acting on some function, thus yielding a full time derivative after hitting the resulting expression with the spinor derivative coming
from the measure. Such a condition is slightly
more complicated for analysis that the one which follows after integration over $\theta$.}
over $\theta$ in \p{S2} and then to replace
the arguments of these functions $\nabla_\theta \eta, \nabla_\theta \lambda$ by
\be\label{ss1}
\nabla_\theta \eta = \left( 1+\eta \dot \eta\right) D\eta \equiv \left( 1+\eta \dot \eta\right) x, \quad
\nabla_\theta \lambda = D\lambda +\eta \dot \lambda D\eta \equiv y + \eta \dot \lambda x.
\ee
Therefore
\be\label{ss2}
F[\nabla_\theta \eta, \nabla_\theta \lambda] = F[x,y] + \eta\dot \eta x F[x,y]_x + \eta \dot \lambda x F[x,y]_y,
\ee
where
\be\label{ss3}
x \equiv D\eta, \qquad y \equiv D\lambda.
\ee
Performing the integration in \p{S2}, expanding the functions $F_1,F_3,F_4$ as in \p{ss2} and collecting the terms
which are linear and cubic in the fermions we will get three terms
\be\label{eq1}
\dot \eta \left[ x F_1 + \left( 1+x^2\right) \left(F_1\right)_x - y F_3\right] - \eta \frac{d}{dt} \left[ x F_1\right] ,
\ee
\be\label{eq2} \dot \lambda \left[ \left( 1+x^2\right) \left(F_1\right)_y +y F_4 \right] -
\lambda \left( \dot {y} F_4 - \dot {x} F_3\right) ,
\ee
\be\label{eq3}
\left(\left( F_3\right)_y +\left(F_4\right)_x \right)\left( \lambda\dot \lambda \dot \eta+
x \left( x \lambda\dot \lambda \dot \eta -\eta\lambda \dot \lambda \dot {x}
+\eta\dot \eta\dot \lambda y - \eta\dot \eta \lambda \dot {y}\right) \right),
\ee
which have to be equal to zero after integration over $t$, independently. The simplest way to get the corresponding equations
is to consider the equations of motion for $\eta$ and $\lambda$ which follow from \p{eq1} - \p{eq3}. Thus, we have three equations
\bea\label{eq4}
\left[ \left(1+x^2\right) F_1 \right]_x - y F_3 =a, && \qquad (a)  \nn\\
y \left[ \left( 1+x^2\right) F_1 \right]_{yy} +\left[ y^2 F_4 \right]_y =0, && \qquad (b) \\
\left(F_3\right)_y +\left(F_4\right)_x=0, && \qquad (c) \nn
\eea
where $a$ in the first equation is an arbitrary constant.

The equation $(\ref{eq4}b)$ may be integrated once, giving
\be\label{eq5}
y \left( \tF_1\right)_y - \tF_1 +y^2 F_4 = G[x],
\ee
where
\be\label{eq6}
\tF_1 \equiv \left( 1+ x^2\right) F_1,
\ee
and $G[x]$ is an arbitrary function depending on $x$ only. Note, that the equation $(\ref{eq4}a)$ now reads
\be\label{eq7}
\left( \tF_1\right)_x - y F_3 =a.
\ee
Differentiating the equation \p{eq5} over $x$ and the equation \p{eq7} over $y$, and using the equation  $(\ref{eq4}c)$
one may get
\be\label{eq8}
G'[x] = -a \qquad G[x]= -a x -b,
\ee
where $b$ is a new constant.

Finally, representing the function $\tF_1$ as
\be\label{eq9}
\tF_1 = a x + b + y \hF_1
\ee
we will finish with the equations
\be\label{eq10}
\left( \hF_1\right)_y+ F_4=0, \qquad \left(\hF_1\right)_x - F_3=0,
\ee
which define the functions $F_3, F_4$ in terms of an arbitrary function $\hF_1$.

Thus, the action \p{S1}, with the functions
\be\label{eq11}
F_1 = \frac{ax + b}{1+x^2} +\frac{y}{1+x^2} \hF_1,\quad F_3 = \left(\hF_1\right)_x,\quad
F_4 = -\left( \hF_1\right)_y
\ee
and arbitrary functions $\hF_1$ and $F_2$, is invariant under the broken $S$ supersymmetry.
The last step is to note that the action \p{S1} with $F_1= \frac{b}{1+x^2}$ is trivial, because
\be\label{triv}
S_{triv} = \int dt d\theta \left( 1 +\eta\dot \eta\right) \frac{\eta}{1+ \left(\nabla_\theta \eta\right)^2} =
\int dt \partial_t q =0
\ee
in view of \p{ih}. Thus, the constant $b$ is unessential and the final result for the functions in the action \p{S1} reads
\be\label{fineq}
F_1 = a \frac{\nabla_\theta \eta}{1+\left(\nabla_\theta \eta\right)^2} +\frac{\nabla_\theta \lambda}{1+\left(\nabla_\theta \eta\right)^2} \hF_1,\quad
F_3 = \left(\hF_1\right)_{\nabla_\theta \eta},\quad
F_4 = -\left( \hF_1\right)_{\nabla_\theta \lambda} ,
\ee
with $\hF_1$ and $F_2$ being arbitrary functions on $\nabla_\theta \eta$ and $\nabla_\theta \lambda$
\be\label{arF}
\hF_1 = \hF_1 [\nabla_\theta \eta, \nabla_\theta \lambda], \qquad F_2 = F_2 [\nabla_\theta \eta, \nabla_\theta \lambda].
\ee
\setcounter{equation}{0}
\section{Interesting cases}
Having at hands the most general expression for the action invariant with respect to both $Q$ and $S$ supersymmetries \p{S1}, \p{fineq},
it is interesting to visualize the functions $F_1,F_2$ which being substituted into \p{S1} will reproduce the FPS actions \p{action-tilde1}, \p{action-tilde}.
Comparing \p{fin1} and \p{fin2} we may conclude, that these two FPS cases are related through the substitutions $\lambda \rightarrow - \im \lambda, y \rightarrow -\im y$.
Therefore, it is enough to consider the first case, with $d_{111}=d_{222}=1$, only. To simplify everything, it is useful to represent the integrand in \p{S1} as follows
\be\label{res1}
L= \eta F_1+ \lambda F_2 +\lambda \eta \dot \eta \left( x F_2 +\hF_1\right)_x-
\eta\lambda \dot \lambda \left( x F_2 + \hF_1 \right)_y,
\ee
with
\be
F_1 \equiv a \frac{x}{1+x^2}+\frac{y}{1+x^2}\hF_1,
\ee
where the functions $\hF_1 $ and $F_2$ depend on $x,y$ variables \p{ss3}. Comparing the integrands in the actions \p{action-tilde1} and \p{res1}, we will get
\bea
L_{FPS} = \nu_1 & \Rightarrow & F_1=F_2= \frac{x+y}{1+(x+y)^2},\;\; a=1, \label{ac1} \\
L_{FPS} = \psi_1 D\psi_2 + \nu_1 D \nu_2 & \Rightarrow & F_1=F_2= \frac{(x-y)(1+x^2-y^2)}{(1+(x-y)^2)(1+(x+y)^2)}, \;\; a=1. \label{ac2}
\eea
Unfortunately, the explicit form of the functions $F_1,F_2$ which corresponds to FPS-like actions \p{ac1}, \p{ac2} is not informative enough to understand
why these actions were selected. The action \p{S1}, with the restrictions \p{fineq}, contains two arbitrary functions $F_1$ and $F_2$. So, as we expected,
the invariance with respect to additional, spontaneously broken $N=1$ supersymmetry does not fix the action in the many particles case, in contrast with the
one particle case, which corresponds to the conditions $\hF_1=F_2=0$ in the superfield Lagrangian \p{res1}.
The bosonic core of the action \p{S1} with the Lagrangian \p{res1} has the form
\be
S_{bos} = \int dt \; \frac{1}{1+\sqrt{1-{\dot q}{}^2}} \left( {\dot q} \cF_1 + {\dot \phi} \cF_2\right),
\ee
where $\cF_1$ and $\cF_2$ are arbitrary functions depending on $\dot q$ and $\dot \phi$.

The explicit form of the Lagrangian in \p{res1} suggests one special case with
\be
\hF_1=-x\,F_2
\ee
for which the terms cubic in the fermions disappear. With such a choice, the bosonic core simplifies to be
\be\label{sss}
{\widehat S}_{bos} = \int dt \, \left( \frac{a}{2}\left(1-\sqrt{1-{\dot q}{}^2}\right) +\frac{1}{2} \dot \phi  \cF_2\right).
\ee
Now, choosing the function $\cF_2$ to be
\be
\cF_2= \frac{b \dot\phi}{1+\sqrt{1-{\dot \phi}{}^2}}
\ee
we will get\footnote{One should remind  that with our definitions \p{ih} and \p{ih2}, the bosonic limits  of the variables
$x,y$ are related with the time derivatives of $q$ and $\phi$ as
$x=\frac{\dot q}{1+\sqrt{1-{\dot q}{}^2}}, y=\frac{\dot \phi}{1+\sqrt{1-{\dot q}{}^2}}$.}
\be
{\widetilde S}_{bos} = \int dt \; \left( \frac{a}{2}\left(1-\sqrt{1-{\dot q}{}^2}\right) + \frac{b}{2}\left(1-\sqrt{1-{\dot \phi}{}^2}\right)\right) .
\ee
This case is indeed special, but again we have no any arguments to clarify this choice.
\setcounter{equation}{0}
\section{Conclusion}
In this paper we analysed in details the application of the FPS approach in one dimension. The basic steps were the same as in the paper \cite{FPS}. The resulting
system is described by the $N=2, d=1$ supersymmetric action for two particles in which one of $N=1$ supersymmetries is spontaneously broken.
The final actions possess the same features as the FPS ones. Using the nonlinear realization approach we reconsider the system in the basis
where only one superfield has the Goldstone nature while the second superfield can be treated as the matter one, being invariant under transformations
of the spontaneously broken $N=1$ supersymmetry. Having at hands
the transformations relating the two selected FPS-like cases with our more generic one, we established the field redefinitions which relate these two cases.
Thus, the only difference between two FPS cases lies in the different choice of the actions, while the supermultiplets specified by the FPS-like constraints
are really the same. In our basis, the two FPS supermultiplets are related by the redefinition of the matter superfield $\lambda \rightarrow -\im \lambda$. Therefore,
it becomes clear why the action with the right signs of the kinetic terms for the bosonic components in one case is mapped, being rewritten in terms
of the second supermultiplet,  into the action with the wrong sign of the kinetic term for one of the bosonic field. Going further with the nonlinear
realization approach, we constructed the most general action for the system of two $N=1$ superfields possessing one additional hidden spontaneously
broken $N=1$ supersymmetry. The constructed action contains two arbitrary functions and is reduced to the FPS actions upon specification of these functions.
Unfortunately, the exact form of these functions corresponding to FPS actions is not very informative and it gives no reason why the FPS cases were selected.

Of course, our consideration was strictly one-dimensional and therefore we cannot argue that all features we discussed will appear in the generalized
supersymmetric Born-Infeld theory constructed in \cite{FPS,FPS2}. Nevertheless, we believe that our results and the generality of the nonlinear
realization approach are quite reasonable tools to reconsider the questions:
\begin{itemize}
 \item whether the supermultiplets with different constants $d_{abc}$ are really different?
 \item which additional properties select the FPS-like actions?
 \end{itemize}
in four dimensions.


\begin{thebibliography}{99}
\bibitem{FPS} S.~Ferrara, M.~Porrati, A.~Sagnotti, {\it N=2 Born-Infeld Attractors}, JHEP {\bf 1412} (2014) 065,\\ {\tt arXiv:1411.4954[hep-th]}.
\bibitem{BG} J.~Bagger, A.~Galperin, {\it New Goldstone multiplet for partially broken supersymmetry}, \\ Phys. Rev. D {\bf 55} (1997) 1091, {\tt arXiv:hepth/9608177}.
\bibitem{FPS2} S.~Ferrara, M.~Porrati, A.~Sagnotti, R.~Stora, A.~Yeranyan, { \it Generalized Born--Infeld Actions and Projective Cubic Curves}, {\tt arXiv:1412.3337 [hep-th]}.
\bibitem{NLR1}S.R.~Coleman, J.~Wess, B.~Zumino,
{\it Structure of phenomenological Lagrangians. 1},
Phys.Rev. 177 (1969) 2239,\\
C.~Callan, S.R.~Coleman, J.~Wess, B.~Zumino,
{\it Structure of phenomenological Lagrangians. 2},
Phys.Rev. 177 (1969) 2247.
\bibitem{NLR2} D.V.~Volkov,
{\it Phenomenological Lagrangians},
Sov.J.Part.Nucl. {\bf 4}(1973) 3;\\
V.I.~Ogievetsky,
{\it Nonlinear realizations of internal and space-time symmetries},
In Proceedings of the Xth Winter School of Theoretical Physics in Karpacz, Vol.1, p.117, 1974.
\bibitem{ikap1} E.A. Ivanov, A.A. Kapustnikov, {\it General Relationship Between Linear and Nonlinear Realizations of Supersymmetry}, J. Phys. A {\bf 11}
(1978) 2375; {\it The Nonlinear Realization Structure Of Models With Spontaneously Broken Supersymmetry }, J. Phys. G {\bf 8} (1982) 167.
\bibitem{comments} S.~Bellucci, S.~Krivonos, A.~Sutulin, {\it Comments on N=2 Born-Infeld Attractors},
{\tt arXiv:1411.5592 [hep-th]}.
\bibitem{DIK} F.~Delduc, E.~Ivanov, S.~Krivonos, {\it 1/4 Partial Breaking of Global Supersymmetry and New Superparticle Actions},
Nucl. Phys. {\bf B 576} (2000) 196, {\tt arXiv:hep-th/9912222}.
\bibitem{ih} E.A.~Ivanov, V.I.~Ogievetsky, {\it The Inverse Higgs Phenomenon in Nonlinear Realizations},
Teor. Mat. Fiz. {\bf 25} (1975) 164.

\end{thebibliography}
\end{document}